\documentclass{article}
\usepackage{spconf,amsmath,graphicx}
\usepackage{amsmath,graphicx,bbm}
\usepackage{multirow, hhline}
\DeclareMathOperator*{\argmax}{arg\,max}

\title{Domain Adaptation via Teacher-Student Learning for End-to-End Speech Recognition}
\name{Zhong Meng, Jinyu Li, Yashesh Gaur, Yifan Gong}
\address{Microsoft Corporation, Redmond, WA, USA}

\begin{document}
\maketitle

\begin{abstract}
Teacher-student (T/S) has shown to be effective for domain adaptation of deep neural network acoustic models in hybrid speech recognition systems. In this work, we extend the T/S learning to large-scale unsupervised domain adaptation of an attention-based end-to-end (E2E) model through two levels of knowledge transfer: teacher's token posteriors as soft labels and one-best predictions as decoder guidance.
To further improve T/S learning with the help of ground-truth labels, we propose  adaptive T/S (AT/S) learning. Instead of conditionally choosing from either the teacher's soft token posteriors or the one-hot ground-truth label, 
in AT/S, the student always learns from both the teacher and the ground truth with a pair of adaptive weights assigned to the soft and one-hot labels quantifying the confidence on each of the knowledge sources. The confidence scores are dynamically estimated at each decoder step as a function of the soft and one-hot labels. With 3400 hours parallel close-talk and far-field Microsoft Cortana data for domain adaptation, T/S and AT/S achieve 6.3\% and 10.3\% relative word error rate improvement over a strong E2E model trained with the same amount of far-field data.
\end{abstract}


\section{Introduction}
Recently, with the advancement of deep learning, great progress has been made in end-to-end (E2E) automatic speech recognition (ASR). With the goal of directly mapping a sequence of speech frames to a sequence of output tokens, an E2E ASR system incorporates the acoustic model,  language model and  pronunciation model of a conventional  ASR system into a single deep neural network (DNN). The most dominant approaches for E2E ASR include connectionist temporal classification (CTC) \cite{graves2006connectionist, graves2014towards}, recurrent neural network transducer (RNNT) \cite{graves2012sequence} and attention-based encoder-decoder (AED) models \cite{chorowski2015attention, bahdanau2016end, chan2016listen}.

However, the performance of E2E ASR degrades significantly when an acoustic mismatch exists between training and test conditions. An intuitive solution is domain adaptation where a well-trained source-domain E2E model is adapted to the data in the target domain. Different from speaker adaption, domain adaptation allows for the usage of a large amount of adaptation data in both source and target domains.

There has been plenty of domain adaptation methods for hybrid systems that we can leverage for adapting E2E systems.  One popular approach is the adversarial learning  in which an intermediate deep feature \cite{grl_sun, dsn_meng, meng2019asa} or a front-end speech feature \cite{meng2018cycle, meng2018afm} is learned to be 
invariant to the shifts between source and target domains. 
Adversarial domain adaptation is suitable for the situation where no transcription or \emph{parallel} adaptation data in both domains are available. It can also effectively suppress the environment \cite{meng2018adversarial, grl_shinohara,meng2019aadit} and speaker \cite{meng2018speaker, saon2017english} variability during domain adaptation.
However, in speech area, a parallel sequence of target-domain data can be easily simulated from the source-domain data such that the speech from both domains are frame-by-frame synchronized. To take advantage of this, teacher-student (T/S) learning \cite{li2014learning} was proposed for the \emph{unsupervised} domain adaptation of acoustic models in DNN-hidden Markov model (HMM) hybrid systems \cite{li2017large}.
In T/S learning, the Kullback-Leibler (KL) divergence between the output senone distributions of teacher and student acoustic models given parallel source and target domain data at the input is minimized by updating only the student model parameters. T/S training was shown to outperform the cross entropy training directly using the hard label in the target domain \cite{li2017large, Watanabe17, Li2018Speaker, meng2019conditional, asami2017domain}. 

One drawback of unsupervised T/S learning is that, the teacher model is not perfect and will sometimes make inaccurate predictions that mislead the student model toward suboptimal directions. To overcome this, one-hot ground-truth labels are used to compensate for teacher's imperfections.  Hinton et al. proposed interpolated T/S (IT/S) learning \cite{hinton2015distilling} to interpolate the  teacher's soft class posteriors with one-hot ground truth using a pair of globally fixed weights. However, the optimal weights are data-dependent and can only be determined through careful tuning on a dev set. More recently, conditional T/S (CT/S) learning was proposed in \cite{meng2019conditional} where the student model selectively chooses to learn from either the teacher or the ground truth depending on whether the teacher's prediction is correct or not. CT/S does not disturb the statistical relationships among classes naturally embedded in the class posteriors and achieves significant word error rate (WER) improvement over T/S for domain adaptation on CHiME-3 dataset \cite{chime3_barker}.

In this work, we focus on the domain adaptation of AED models for E2E ASR by using T/S learning which was previously applied to learn small-footprint AED models in \cite{kim2016sequence, mun2019sequence, pang2018compression} by distilling knowledge from a large powerful teacher AED. 
For unsupervised domain adaptation, we extend T/S learning to AED models by introducing a two-level knowledge transfer: in addition to learning from the teacher's soft token posteriors, the student AED also conditions its decoder on the one-best token sequence decoded by the teacher AED. 

We further propose an \emph{adaptive T/S (AT/S) learning} method to improve T/S learning using ground-truth labels. By taking advantage of both IT/S and CT/S, AT/S \emph{adaptively} assigns a pair of weights to the teacher's soft token posteriors and the one-hot ground-truth label at each decoder step depending on the confidence scores on each of the labels. The confidence scores are dynamically estimated as a function of soft and one-hot labels. The student AED learns from an adaptive linear combination of both labels. AT/S inherits the linear interpolation of soft and one-hot labels from IT/S and borrows from CT/S the judgement on the credibility of both knowledge sources before merging them. It is expected to achieve improved performance over the other T/S methods for domain adaptation. As a general deep learning method, AT/S can be widely applied to the domain adaptation or model compression of any DNN. 

With 3400 hours close-talk and far-field Microsoft Cortana data for domain adaptation, T/S learning achieves up to 24.9\% and 6.3\% relative WER gains over close-talk and far-field baseline AEDs, respectively. AT/S improves the close-talk and far-field AEDs by 28.2\% and 10.3\%, respectively, consistently outperforming IT/S and CT/S.

\section{Attention-Based Encoder-Decoder (AED) Model}
In this work, we perform domain adaptation on AED
models \cite{chorowski2015attention, bahdanau2016end, chan2016listen}. AED model was first introduced in
\cite{cho2014properties, bahdanau2014neural} for neural machine
translation. Without any conditional independence assumption
as in CTC \cite{graves2006connectionist}, AED was successfully applied to to E2E ASR in \cite{chorowski2015attention, bahdanau2016end, chan2016listen} and has recently achieved superior performance to conventional hybrid systems in \cite{chiu2018state}. 

AED directly models the conditional probability distribution
$P(\mathbf{Y} | \mathbf{X})$ over sequences of output tokens
$\mathbf{Y}=\{y_1, \ldots, y_L\}$ given a sequence of input speech frames
$\mathbf{X}=\{\mathbf{x}_1, \ldots, \mathbf{x}_N\}$ as below:
\begin{align}
	P(\mathbf{Y}|\mathbf{X}) = \prod_{l=1}^L P(y_l | \mathbf{Y}_{0:l-1},
	\mathbf{X}). 
\end{align}

To achieve this, the AED model incorporates an encoder, a decoder and an
attention network. 
The encoder maps a sequence of input speech frames $\mathbf{X}$
into a sequence of high-level features $\mathbf{H} =
\{\mathbf{h}_1, \ldots, \mathbf{h}_N\}$ through an RNN.
An attention network is used to 
determine
which encoded features in $\mathbf{H}$ should be attended to predict
the output label $y_l$ and to generate a context vector $\mathbf{z}_l$ as a linear combination of $\mathbf{H}$ \cite{chorowski2015attention}.
A decoder is used to model $P(\mathbf{Y}|\mathbf{H})$ which is equivalent to $P(\mathbf{Y}|\mathbf{X})$.  At each
time step $t$, the decoder RNN takes the sum of the previous token embedding 
$\mathbf{e}_{l-1}$ and the context vector $\mathbf{z}_{l-1}$ as the
input to predict the conditional probability of each token, i.e., $P(u |
\mathbf{Y}_{0:l-1}, \mathbf{H}), u \in \mathbbm{U}$, at the decoder step $l$, where
$\mathbbm{U}$ is the set of all the output tokens:
\begin{align}
        \mathbf{q}_l &= \text{RNN}^{\text{dec}}(\mathbf{q}_{l-1}, \mathbf{e}_{l-1} + \mathbf{z}_{l-1}), \label{eqn:decoder_rnn} \footnotemark \\
        \left[P(u | \mathbf{Y}_{0:l-1}, \mathbf{X})\right]_{u \in
       \mathbbm{U}} &= \text{softmax}\left[K_{y}(\mathbf{q}_l + \mathbf{z}_l) +
       \mathbf{b}_y\right], \label{eqn:decoder_fc}
\end{align}
\footnotetext{In Eq.~\eqref{eqn:decoder_rnn} and Eq.~\eqref{eqn:decoder_fc}, we sum together the $\mathbf{z}_l$ and $\mathbf{q}_l$ (or $\mathbf{e}_t$) instead of concatenation, because, by summation, we get a lower-dimensional combined vector than concatenation, saving the number of parameters by half for the subsequent projection operation. In our experiments, concatenation does not improve the performance even with more parameters.}
where $\mathbf{q}_l$ is the hidden state of the decoder RNN. bias $\mathbf{b}_y$ and the matrix $K_y$ are learnable parameters. 

An AED model is trained to minimize the following cross-entropy (CE) loss on the training corpus $\mathbbm{T_r}$.
\begin{align}
	\mathcal{L}_{\text{CE}} (\theta, \mathbbm{T_r}) 
	& =-\hspace{-10pt}\sum_{(\mathbf{X}, \mathbf{Y}^G) \in \mathbbm{T_r}} \sum_{l = 1}^{L^G} \log P(y^G_l |
		\mathbf{Y}^G_{0:l-1}, \mathbf{X}; \theta)
       \label{eqn:ce}
\end{align}
where $\mathbf{Y}^G = \{y^G_1, \ldots, y^G_{L^G}\}$ is the sequence of grouth-truth tokens, $L^G$ represents the number of elements in $\mathbf{Y}^G$ and $\theta$ denotes all the model parameters in AED.

\section{T/S Learning for Unsupervised Domain Adaptation of AED}
\label{sec:ts}

\begin{figure}[htpb!]
	\centering
	\includegraphics[width=1.0\columnwidth]{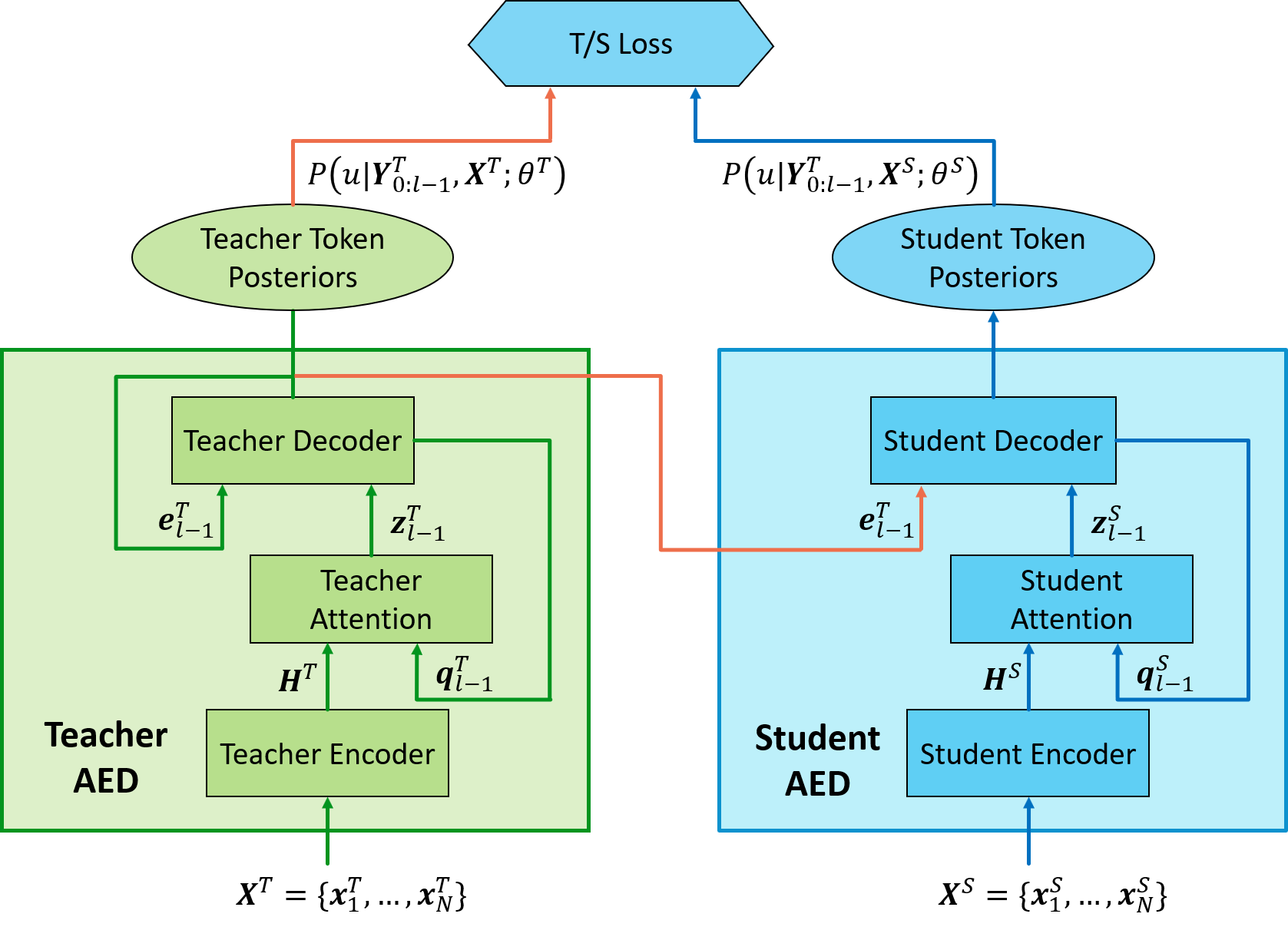}
\vspace{-5pt}
	\caption{T/S learning for unsupervised domain adaptation of AED model for E2E ASR. The two orange lines signify the two-level knowledge transfer.}
	\label{fig:ts}
\end{figure}

For unsupervised domain adaptation, we want to make use of a large amount of \emph{unlabeled} data that is widely available. As shown in Fig. \ref{fig:ts}, with T/S learning, only two sequences of parallel data are required: an input sequence of source-domain speech frames to the teacher AED 
$\mathbf{X}^T=\{\mathbf{x}^T_{1}, \ldots, \mathbf{x}^T_{N}\}$ and an input sequence of target-domain speech frames to the student model $\mathbf{X}^S=\{\mathbf{x}^S_{1}, \ldots, \mathbf{x}^S_{N}\}$.  $\mathbf{X}^T$ and
$\mathbf{X}^S$ are parallel to each other, i.e, each pair of $\mathbf{x}^S_n$ and $\mathbf{x}^T_n, \forall n \in \{1, \ldots, N\}$ are frame-by-frame synchronized. For most domain adaptation tasks in ASR, such as adapting from clean to noisy speech, close-talk to far-field speech, wide-band to narrow-band speech, the parallel data in the target domain can be easily simulated from the data in the source domain \cite{li2017large, Li2018Speaker}. 

Our goal is to train a student AED that can accurately predict the tokens of the target-domain data by forcing the student to emulate the behaviors of the teacher. To achieve this, we minimize the Kullback-Leibler (KL) divergence between the \emph{token-level} output distributions of the teacher and the student AEDs given the parrallel data $\mathbf{X}^T$ and $\mathbf{X}^S$ are fed as the input to the AEDs. The KL divergence between the token-level output distributions of the teacher and student AEDs are formulated below
\begin{align}
&\hspace{-7pt} \sum_{l=1}^{L^T}\sum_{u \in \mathbbm{U}}
P(u|\mathbf{Y}^{T}_{0:l-1}, \mathbf{X}^{T}; \mathbf{\theta}^{T}) \log
\left[ \frac{P(u|\mathbf{Y}^{T}_{0:l-1},
	\mathbf{X}^{T};\mathbf{\theta}^{T})}{P(u|\mathbf{Y}^{T}_{0:l-1},
		\mathbf{X}^{S}; \mathbf{\theta}^{S})}
	\right],
	\label{eqn:kld}
\end{align}
where $\mathbf{Y}^{T}=\{y^T_1, \ldots, y^T_{L^T}\}$ is the sequence of \emph{one-best} token sequence decoded by the teacher AED as follows  
\begin{align}
    y^T_l = \argmax_{u \in \mathbbm{U}}P(u | \mathbf{Y}^T_{0:l-1}, \mathbf{X}^T), \quad l = 1, \ldots, L^T,
	\label{eqn:one_best}
\end{align}
where $L^T$ is the number of tokens in $\mathbf{Y}^T$, and $\mathbf{\theta}^{T}$, $\mathbf{\theta}^{S}$ denote all the parameters in the teacher and student AED models, respectively. Note that, for unsupervised domain adaptation, the teacher AED can only condition its decoder on the token $y^T_{l-1}$ predicted at the previous step since the ground-truth labels $\mathbf{Y}^G$ are not available. We minimize the KL divergence with respect to $\theta^S$ while keeping $\theta^T$ fixed on the adaptation data corpus $\mathbbm{A}$, which is equivalent to minimizing the token-level T/S loss function below:
\begin{align}
& \mathcal{L}_{\text{TS}}(
\mathbf{\theta}^{S}, \mathbbm{A}) = -\sum_{ (\mathbf{X}^T, \mathbf{X}^S) \in \mathbbm{A}} \sum_{l=1}^{L^T}\sum_{u \in \mathbbm{U}}
	P(u|\mathbf{Y}^T_{0:l-1},
	\mathbf{X}^T; \mathbf{\theta}^T) \nonumber \\
	& \qquad \qquad \qquad \quad \qquad \qquad \; \log P(u|\mathbf{Y}^T_{0:l-1},\mathbf{X}^S;
	\mathbf{\theta}^{S}).
	\label{eqn:ts}
\end{align}

The steps of token-level T/S learning for unsupervised domain adaptation of AED model are summarized as follows:
\begin{enumerate}
    \item Clone the student AED from a teacher AED well-trained with transcribed source-domain data by minimizing Eq. \eqref{eqn:ce}.
    \item Forward-propagate the source-domain data $\mathbf{X}^T$ through the teacher AED, generate teacher's one-best token sequence $\mathbf{Y}^T$ using Eq. \eqref{eqn:one_best} and teacher's soft posteriors for each decoder step $P(u|\mathbf{Y}^T_{0:l-1},\mathbf{X}^T; \mathbf{\theta}^{T}), u \in \mathbbm{U}$ by Eqs. \eqref{eqn:decoder_rnn} and \eqref{eqn:decoder_fc}. \label{item:forward}
    \item Forward-propagate the target-domain data $\mathbf{X}^S$ (\emph{parallel} to $\mathbf{X}^T$) through the student AED, generate student's soft posteriors for each teacher's decoder step $P(u|\mathbf{Y}^T_{0:l-1},$ $\mathbf{X}^S; \mathbf{\theta}^{S}), u \in \mathbbm{U}$ by Eqs. \eqref{eqn:decoder_rnn} and \eqref{eqn:decoder_fc}.
    \item Compute error signal of the T/S loss function in Eq. \eqref{eqn:ts} , back-propagate the error through student AED and update the parameters of the student AED. \label{item:backward}
    \item Repeat Steps \ref{item:forward} to \ref{item:backward} until convergence.
\end{enumerate}
After T/S learning, only the adapted student AED is used for testing and the teacher AED is discarded.

From Eqs.~\eqref{eqn:one_best} and \eqref{eqn:ts}, to extend T/S learning to AED-based E2E models, two levels of knowledge transfer are involved: 1) the student learns from the teacher's soft token posteriors $P(u|\mathbf{Y}^T_{0:l-1}, \mathbf{X}^T; \mathbf{\theta}^T)$ at each decoder step; 2) the student AED conditions its decoder on the previous token $y^T_{l-1}$ predicted by the teacher to make the current prediction.


Sequence-level T/S learning \cite{kim2016sequence, wong2016sequence} is another method for unsupervised domain adaptation in which a KL divergence between the \emph{sequence-level} output distributions of the teacher and student AEDs are minimized.
Equivalently, we minimize the sequence-level T/S loss function below with respect to $\theta^S$
\begin{align}
& \mathcal{L}_{\text{TS-SEQ}}(
\mathbf{\theta}^{S}, \mathbbm{A}) \nonumber \\
& = -\sum_{ (\mathbf{X}^T, \mathbf{X}^S) \in \mathbbm{A}} \sum_{\mathbf{V} \in \mathbbm{V}} P(\mathbf{V}| \mathbf{X}^T; \mathbf{\theta}^T)\log P(\mathbf{V}|\mathbf{X}^S;
	\mathbf{\theta}^{S}) \nonumber \\
	& \approx -\sum_{ (\mathbf{X}^T, \mathbf{X}^S) \in \mathbbm{A}} 
	\log P(\mathbf{Y}^T|\mathbf{X}^S;
	\mathbf{\theta}^{S}) \nonumber \\
	& = -\sum_{ (\mathbf{X}^T, \mathbf{X}^S) \in \mathbbm{A}} \sum_{l=1}^{L^T}
	\log P(y^T_l|\mathbf{Y}^T_{0:l-1},\mathbf{X}^S;
	\mathbf{\theta}^{S}),
	\label{eqn:ts_seq}
\end{align}
where $\mathbbm{V}$ is the set of all possible token sequences and the teacher's sequence-level output distribution $P(\mathbf{V} | \mathbf{X}^T)$ is approximated by $\mathbbm{1}[\mathbf{V}=\mathbf{Y}^T]$ for easy implementation. $\mathbbm{1}[\cdot]$ is an indicator function which equals to 1 if the condition in the squared bracket is satisfied and 0 otherwise.

From Eq. \eqref{eqn:ts_seq}, we see that only one level of knowledge transfer exists in sequence-level T/S, i.e., the one-best token sequence $\mathbf{Y}^T$ decoded by the teacher AED. The student AED learns from $\mathbf{Y}^T$ and conditions its decoder on it at each step. Different from token-level T/S, in sequence-level T/S, one-hot labels in $\mathbf{Y}^T$ are used as training targets of the student AED instead of the soft token posteriors.

\section{Adaptive T/S (AT/S) Learning for Supervised Domain Adaptation of AED}
\label{sec:ats}
In this section, we want to make good use of the ground-truth labels of the adaptation data to further improve the T/S domain adaptation. Note that different from unsupervised T/S in Section \ref{sec:ts}, in supervised domain adaptation, the teacher AED conditions its decoder on the ground-truth token instead of its previous decoding result because the token transcription $\mathbf{Y}^G$ is available in addition to $\mathbf{X}^S$ and $\mathbf{X}^T$.

One shortcoming of unsupervised T/S learning is that the teacher model can sporadically predict inaccurate token posteriors which misleads the student AED towards suboptimal performance. One-hot ground-truth labels can be utilized to alleviate this issue. 
One possible solution is the interpolated T/S (IT/S) learning \cite{hinton2015distilling} in which a weighted sum of teacher's soft posteriors and the one-hot ground truth is used as the target to train the student AED. A pair of \emph{global} weights summed to be one is applied to each pair of soft and one-hot labels. However, the optimal global weights are hard to determine because they are data-dependent and need to be carefully tuned on a dev set.

To address this issue, conditional T/S learning (CT/S) \cite{meng2019conditional} was proposed recently in which the student selectively chooses to learn from either the teacher AED or the ground truth conditioned on whether the teacher AED can correctly predict the ground-truth labels. CT/S have shown significant WER improvements over T/S and IT/S for both domain and speaker adaptation on CHiME-3 dataset. However, in CT/S, the student is still not ``smart'' enough because, for each token, the student AED solely relies on either the teacher's posteriors or the ground truth instead of dynamically extracting useful knowledge from both. 

\begin{figure}[htpb!]
	\centering
	\includegraphics[width=1.0\columnwidth]{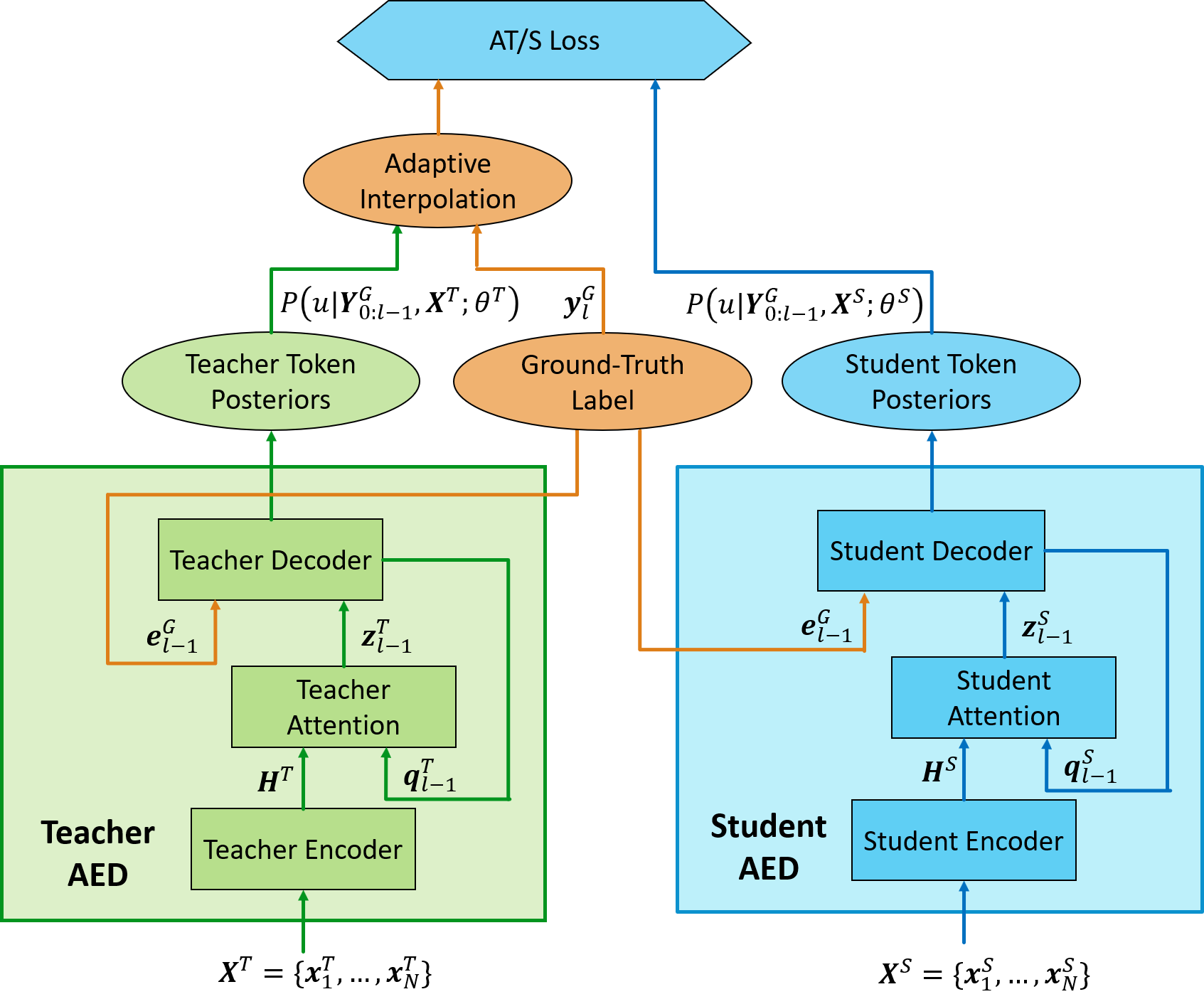}
\vspace{-5pt}
	\caption{Adaptive T/S (AT/S) learning for supervised domain adaptation of AED model for E2E ASR.} 
	\label{fig:ats}
\end{figure}

To further improve the effectiveness of knowledge transfer, we propose an \emph{adaptive teacher-student (AT/S) learning} method by taking advantage of both CT/S and IT/S. 
As shown in Fig. \ref{fig:ats}, instead of assigning a fixed pair of soft weight $w$ and one-hot weight $(1-w)$ for all the decoder steps, we adaptively weight the teacher's soft posteriors at the $l^\text{th}$ decoder step, $P(u|\mathbf{Y}^G_{0:l-1},\mathbf{X}^T;\mathbf{\theta}^{T}), u\in \mathbbm{U}$, by $w_l \in [0,1]$ and the one-hot vector of the $l^\text{th}$ token in the ground-truth sequence $\mathbf{Y}^G$ by $(1-w_l)$.
In order to quantify the value of the knowledge to be transferred, $w_l$ should be positively correlated with a confidence score $c_l$ on the teacher's prediction on token posteriors, while $(1-w_l)$ should be positively correlated with a confidence score on the ground truth $d_l$. To achieve this, we compute $w_l$ by normalizing $c_l$ against its summation with $d_l$.
\begin{align}
    w_l = \frac{c_l}{c_l + d_l}
    \label{eqn:wl}
\end{align}


It is in general true that the higher posterior $P(y_l^G|\mathbf{Y}^G_{0:l-1},$ $\mathbf{X}^T;\theta^T)$ a teacher assigns to the correct (ground-truth) token $y_l^G$, the more accurate the teacher's soft posteriors are at this decoder step. Therefore, the confidence score $c_l$ on teacher's soft posteriors $P(u|\mathbf{Y}^G_{0:l-1}, \mathbf{X}^T; \theta^T), u\in \mathbbm{U}$ can be any \emph{monotonically increasing} function of the correct token posterior predicted by the teacher $P(y^G_l|\mathbf{Y}^G_{0:l-1}, \mathbf{X}^T; \theta^T)$, while the confidence score $d_l$ on the one-hot ground truth can be any monotonically increasing function of $(1-P(y^G_l|\mathbf{Y}^G_{0:l-1}, \mathbf{X}^T;$ $\theta^T))$ as follow
\begin{align}
    c_l &= f_1(P(y^G_l|\mathbf{Y}^G_{0:l-1}, \mathbf{X}^T; \theta^T)), \label{eqn:cl} \\
    d_l &= f_2(1-P(y^G_l|\mathbf{Y}^G_{0:l-1}, \mathbf{X}^T; \theta^T)), \label{eqn:dl}
\end{align}
where both $f_1$ and $f_2$ are any monotonically increasing functions on the interval $[0, 1]$.
In this work, we simply assume that $f_1$ and $f_2$ are both power functions of the same form, i.e., $f_1(x) =  f_2(x) = x^{\lambda}, \; \lambda > 0 $. Note that $w_l$ equals to $P(y^G_l|\mathbf{Y}^G_{0:l-1}, \mathbf{X}^T; \theta^T)$ when $\lambda=1$.


In AT/S, a linear combination of the teacher's soft posteriors and the one-hot ground truth weighted by $w_l$ and $(1- w_l)$, respectively, is used as the training target for the student AED at each decoder step. The AT/S loss function is formulated as
\begin{align}
& \hspace{-7pt} \mathcal{L}_{\text{ATS}}(
\mathbf{\theta}^{S}, \mathbbm{A}) \hspace{-2pt}= \hspace{-1pt} -\hspace{-24pt}\sum_{(\mathbf{X}^T, \mathbf{X}^S, \mathbf{Y}^G) \in \mathbbm{A}} \hspace{-1pt} \sum_{l=1}^{L^G} \hspace{-1pt} \sum_{u \in \mathbbm{U}}
	\left[ w_l P(u|\mathbf{Y}^G_{0:l-1},
	\mathbf{X}^T; \mathbf{\theta}^T) \right. \nonumber \\
	& \left. \vphantom{P(u|\mathbf{Y}^G_{0:l-1},
	\mathbf{X}^T; \mathbf{\theta}^T)} \quad + (1-w_l) \mathbbm{1}[u = y^G_l]\right] \log P(u|\mathbf{Y}^G_{0:l-1},\mathbf{X}^S;
	\mathbf{\theta}^{S}).
	\label{eqn:ats}
\end{align}

The steps of AT/S learning for supervised domain adaptation of AED model are summarized as follows:
\begin{enumerate}
    \item Perform token-level unsupervised T/S adaptation by following the steps in Section \ref{sec:ts} as the initialization.
    \item Forward-propagate the \emph{parallel} source and target domain data $\mathbf{X}^T$ and $\mathbf{X}^S$ through the teacher and student AEDs, generate teacher and student's soft posteriors $P(u|\mathbf{Y}^G_{0:l-1},\mathbf{X}^T; \mathbf{\theta}^{T})$ and $P(u|\mathbf{Y}^G_{0:l-1},$ $\mathbf{X}^S; \mathbf{\theta}^{S}), u \in \mathbbm{U}$ for each decoder step by Eqs. \eqref{eqn:decoder_rnn} and \eqref{eqn:decoder_fc}. \label{item:forward}
    \item Compute the confidence scores $c_l$ and $d_l$ for teacher's soft posteriors and one-hot vector of ground truth $y^G_l$ by Eqs. \eqref{eqn:cl} and \eqref{eqn:dl}, compute the adaptive weight $w_l$ by Eq. \eqref{eqn:wl}.
    \item Compute error signal of the AT/S loss function in Eq. \eqref{eqn:ats} , back-propagate the error through student AED and update the parameters of the student AED. \label{item:backward}
    \item Repeat Steps \ref{item:forward} to \ref{item:backward} until convergence.
\end{enumerate}

AT/S is superior to IT/S in that the combination weights for soft and one-hot labels at each decoder step are adaptively assigned according to the confidence score on both labels. AT/S will degenerate to IT/S if the combination weights $w_l$ are fixed globally. Compared to CT/S, in AT/S, the student always adaptively learns from both the teacher's soft posteriors and the one-hot ground truth 
rather than choosing either of them depending on the correctness of teacher's prediction.

\section{Experiments}
We adapt a close-talk AED model to the far-field data through various T/S learning methods with parallel close-talk and far-field Microsoft Cortana data for E2E ASR. 
\subsection{Data Preparation}
For both training and adaptation, close-talk data consisting of 3400 hours of Microsoft live US English Cortana utterances are collected through a number of deployed speech
services including voice search and SMD. We simulate 3400 hours of far-field Microsoft Cortana data by convolving the close-talk signal with different room impulse responses and adding various environmental noise for both training and adaptation. The 3400 hours far-field data is \emph{parallel} with the 3400 hours close-talk data. We collect 17.5k far-field utterances (about 19 hours) from Harman Kardon (HK) speaker as the test set. 

80-dimensional log Mel filter bank features are extracted from the training, adaptation and test speech every 10 ms over a 25 ms window.  
We stack 3 consecutive frames and stride the stacked frame by 30 ms, to form a sequence of 240-dimensional input speech frames.  We first generate 34k mixed-units consisting of words and multi-letter units as in \cite{li2018advancing} based on the training transcription and then tokenize the training, adaptation transcriptions correspondingly.
We insert a special token \texttt{<space>} between every two adjacent words to indicate the word boundary and add \texttt{<sos>}, \texttt{<eos>} to the beginning and end of each utterance, respectively. 

\subsection{AED Baseline System}
We first train an AED model predicting 34k mixed units with 3400 hours close-talk training data and it ground-truth labels for E2E ASR as in \cite{meng2019character, meng2019speaker, gaur2019acoustic}. The encoder is a bi-directional gated recurrent units (GRU)-recurrent neural network (RNN) \cite{cho2014properties, chung2014empirical} with 6 hidden layers, each with 512 hidden units. We use GRU instead of long short-term memory (LSTM) \cite{erdogan2016multi, meng2017deep} for RNN because it has less parameters and is trained faster than LSTM with no loss of performance. Layer normalization \cite{ba2016layer} is applied for each encoder hidden layer. 
Each mixed unit is represented as a 512-dimensional embedding vector. The decoder is a uni-directional GRU-RNN with 2 hidden layers, each with 512 hidden units. 
The 34k-dimensional output layer of the decoder predicts the posteriors of all the mixed units in the vocabulary.
During training, scheduled sampling \cite{bengio2015scheduled} is applied to the decoder with a sampling probability starting at 0.0 and gradually increasing to 0.4 \cite{chiu2018state}. Dropout \cite{srivastava2014dropout} with a probability of 0.1 is used in both encoder and decoder.
A label-smoothed cross-entropy \cite{chorowski2016towards} loss is minimized during training. Greedy decoding is performed to generate the ASR transcription. We use PyTorch \cite{paszke2017automatic} toolkit for the experiments. Table \ref{table:wer} shows that the close-talk AED model achieves 7.58\% and 17.39\% WERs on a close-talk Cortana test set used in \cite{meng2019speaker} and the far-field HK speaker test set, respectively.

Using the well-trained close-talk AED as the initialization, we then train a far-field AED with 3400 hours far-field data and its ground-truth labels by following the same procedure. When evaluated on the HK speaker test set, the baseline far-field AED achieves 13.93\% WER for ASR as in Table \ref{table:wer}.

\begin{table}[h]
\centering
\begin{tabular}[c]{c|c|c|c}
	\hline
	\hline
	Adaptation & Method & WER & WERR \\
	\hline
	\multirow{1}{*}{\begin{tabular}{@{}c@{}} Direct
		\end{tabular}} & Far-Field CE & 13.93 & - \\
	\hline
	\multirow{2}{*}{\begin{tabular}{@{}c@{}} Unsupervised 
		\end{tabular}} & Token T/S & 13.06 & 6.3 \\
	\hhline{~---}
	& Seq T/S & 14.00 & -0.5\\
	\hline
	\multirow{8}{*}{\begin{tabular}{@{}c@{}} Supervised 
	\end{tabular}} 
	& IT/S ($w = 0.2$) & 12.95 & 7.0 \\
	\hhline{~---}
	& IT/S ($w = 0.8$) & 12.96 & 7.0 \\
	\hhline{~---}
	& CT/S & 12.82 & 8.0 \\
	\hhline{~---}
	& AT/S ($\lambda = 0.10$) & 12.56 & 9.8 \\
	\hhline{~---}
	& AT/S ($\lambda = 0.25$) & \textbf{12.49} & \textbf{10.3} \\
	\hhline{~---}
	& AT/S ($\lambda = 1.0$) & 12.66 & 9.1 \\
	\hhline{~---}
	& AT/S ($\lambda = 3.0$) & 12.71 & 8.8 \\
	\hline
	\hline
	\end{tabular}
  \caption{The ASR WER (\%) of far-field AEDs trained with CE and AED models adapted by various T/S learning methods to 3400 hours far-field Microsoft Cortana data for E2E ASR on HK speaker test set. ``Seq T/S'' stands for sequence-level T/S and WERR (\%) represents relative WER reduction.}
\label{table:wer}
\vspace{-15pt}
\end{table}

\subsection{Unsupervised Domain Adaptation with T/S Learning}
We adapt the close-talk baseline AED to the 3400 hours far-field data using token and sequence level T/S learning as discussed in Section \ref{sec:ts}. To achieve this, we feed the 3400 hours close-talk adaptation data as the input to the teacher AED and the 3400 hours \emph{parallel} far-field adaptation data as the input to the student AED. The student AED conditions its decoder on one-best token sequences generated by the teacher AED through greedy decoding. In token-level T/S, the soft posteriors generated by the teacher serve as the training targets of the student while in sequence-level T/S, the one-best sequences decoded by the teacher are used the targets.

As shown in Table \ref{table:wer}, the token-level T/S achieves 13.06\% WER on HK speaker test set, which is 24.9\% and 6.25\% relative improvements over the close-talk and far-field AED models, respectively. The sequence-level T/S achieves 14.00\% WER, which is 19.5\% relative improvement over the close-talk AED model. The sequence-level T/S performs slightly worse than the far-field AED trained with ground-truth labels because the one-best decoding from the teacher AED is not always reliable to serve as the training targets for the student model. 
The sequence-level T/S can be improved by using multiple decoded hypotheses generated by the teacher AED as the training targets as in \cite{mun2019sequence, pang2018compression}. We did not perform N-best decoding because it will drastically increase the computational cost and will consumes much more adaptation time than the other T/S methods. 
The 6.7\% relative WER gain obtained by token-level T/S over sequence-level T/S shows the benefit of using soft posteriors generated by the teacher AED as the training target at each decoder step when a reliable ground-truth transcription is not available.  

The 6.3\% relative WER gain of token T/S over far-field AED baseline shows that the unsupervised T/S learning with no ground-truth labels can significantly outperform the supervised domain adaptation with such information available. Compared to the one-hot labels, the soft posteriors accurately models the inherent statistical relationships among different token classes in addition to the token identity encoded by a one-hot vector. It proves to be a more powerful target for the student to learn from which is consistent with what was observed in \cite{li2017large, Watanabe17, Li2018Speaker, meng2019conditional, asami2017domain}.

\subsection{Supervised Domain Adaptation with AT/S Learning}
As discussed in Section \ref{sec:ats}, we want to further improve the T/S learning by using one-hot ground-truth labels when they are available. As in \cite{hinton2015distilling}, we perform IT/S learning for supervised domain adaptation by using the linear interpolation of soft posterior and one-hot ground truth as the training target of the student. The interpolation weights are globally fixed at 0.5 and 0.5 for all decoder steps. By following \cite{meng2019conditional}, we also conduct CT/S for supervised domain adaptation where soft posteriors are used as the training target of the student if the teacher's prediction is correct at the current decoder step, otherwise the one-hot ground truth is used as the target. Finally, AT/S domain adaptation is performed by adaptively adjusting the weights assigned to the soft and one-hot labels at each decoder step as in Eqs. \eqref{eqn:wl} to \eqref{eqn:dl}. We explore using different power functions as $f_1(x)$ and $f_2(x)$ to compute the confidence scores by adjusting $\lambda$. For all the above supervised T/S learning methods, the 3400 hours close-talk and 3400 hours far-field parallel adaptation data is fed as the input to the teacher and student AEDs, respectively.

As shown in Table \ref{table:wer}, IT/S  with $w = 0.2$ achieves 13.95\% WER on HK speaker test set which is 25.5\%, 7.0\% and 0.8\% relative improvements over the close-talk, far-field and token-level T/S adapted AED models, respectively. With a 12.82\% WER, CT/S relatively improves the close-talk, far-field and token-level T/S adapted AED models by 26.3\%, 8.0\% and 1.8\% respectively. Among different $\lambda$s for AT/S, the best WER is 12.49\%, which is 28.2\%, 10.3\% and 4.4\% relative gains over close-talk, far-field and token-level T/S adapted AEDs. The minimum WER is reached when $\lambda=0.25$ and $c_l = P(y_l|\mathbf{Y}^G_{0:l-1}, \mathbf{X}^T; \theta^T)^{0.25}$. Compared to $\lambda>1$, AT/S works better for $\lambda \in[0, 1]$ when confidence scores $c_l$, $d_l$ are both concave functions of the correct token posterior and the sum of incorrect token posteriors, respectively. All the IT/S, CT/S and AT/S outperform the unsupervised T/S learning indicating that the one-hot ground truth can further improve T/S domain adaptation when it is properly used. AT/S achieves the largest gain in supervised domain adaptation methods showing the superiority of adaptively extracting useful knowledge from both the soft and one-hot labels depending on their confidence scores.

\section{Conclusion}
In this paper, we extend T/S learning to unsupervised domain adaptation of AED models for E2E ASR. T/S learning requires only unlabeled parallel source and target domain data as the input to the teacher and student AEDs, respectively. In T/S, the student AED conditions its decoder on the one-best token sequences generated by the teacher. The teacher's soft posteriors and decoded one-hot tokens are used as the training target of the student AED for token-level and sequence-level T/S learning, respectively.

For supervised domain adaption, we propose adaptive T/S learning in which the student always learns from a linear combination of the teacher's soft posteriors and the one-hot ground truth. The combination weights are adaptively computed at each decoder step based on the confidence scores on both knowledge sources.

Domain adaptation is conducted on 3400 hours close-talk and 3400 hours far-field Microsoft Cortana data. Token-level T/S achieves 6.3\% relative WER improvement over the baseline far-field AED model trained with CE criterion. By making use of the ground-truth labels, AT/S further improves the token-level T/S by 4.4\% relative and achieves a total 10.3\% relative gain over the far-field AED. AT/S also consistently outperforms IT/S and CT/S showing the advantage of learning from both the teacher and the ground truth as well as  the adaptive adjustment of the combination weights. 


\vfill\pagebreak

\bibliographystyle{IEEEtran}
\bibliography{refs}

\end{document}